\documentclass[journal=jpcafh,manuscript=article,layout=twocolumn]{achemso}
\usepackage{aas_macros}
\usepackage{natbib}
\usepackage{amsmath}
\usepackage{wasysym}
\usepackage{graphicx}
\usepackage{calc}
\usepackage{hyperref}
\usepackage{placeins}
\usepackage{caption}

\usepackage{chemformula} 
\usepackage[T1]{fontenc} 

\author{Tyler Nelson}
\email{tyler.neslon@utexas.edu}
\author{Anita L. Cochran}
\email{anita@astro.as.utexas.edu}
\altaffiliation{McDonald Observatory}
\affiliation{University of Texas At Austin}
\author{Colin Western}
\affiliation{University of Bristol}
\title{Rotational Temperature Modeling of the Swan $\Delta\nu = 0$ Band Sequence in Comet 122P/de~Vico}
\begin{document}
	
	\begin{abstract}
		We modeled observations of the C$_2$ $\mathrm{d} ^3\Pi_g - \mathrm{a} ^3\Pi_u$ (Swan) $\Delta \nu = 0$ sequence observed in spectra of comet 122P/de~Vico obtained with the 2.7\,m Harlan J. Smith Telescope and Tull Coude spectrograph of McDonald observatory on 10/03/1995 and 10/04/1995. The data used spanned 4986-5169\,\AA \ at R\,=\,$\lambda/\Delta\lambda$\,=\,60,000. We used the PGOPHER molecular spectra model
		to generate and fit synthetic spectra with the d$^3\Pi_g$ having one and two rotational temperatures. We found the excited state had a two component rotational temperature, similar to that found for comet Halley. The modeled spectrum was sufficiently high quality that local perturbations were important to include. The large perturbation, b$^3\Sigma_g^-(\nu=10)$, was added to our fits and some new estimates on its molecular constants were found.
	\end{abstract}
	
	\section{Introduction}
	
	Comets and asteroids represent the leftovers from the epoch of planet formation within the solar system. Comets differ from the rocky asteroids by the presence of ices, which are a direct indication that they formed in a region beyond Jupiter. Since comets spend most of their orbit in cold regions and have insufficient mass to differentiate or undergo structure-altering processes, they contain the least altered material from the primordial solar system. Thus, studies of comets offer constraints on conditions in the early solar nebula.
	
	Comets are mixtures of ices and dust, with the mass being split approximately
	equally between those two substances. The cometary ice is about 80\% H$_2$O,
	the remainder having contributions from other ices such as CO, CO$_2$, CH$_4$, NH$_3$, etc.
	As the comet approaches the Sun, the ices are sublimated and flow out
	from the nucleus carrying dust along with it. The dust quickly disentrains from
	the gas, flowing outwards more slowly than the gas.  The gas travels
	outwards from the nucleus at a velocity that is dependent on heliocentric
	distance, being approximately 0.85 km\,sec$^{-1}$ at 1\,{\sc au}.
	Beyond a short distance from the nucleus (500 -- 1000\,km depending on how
	active the comet is) the density of the gas is too low for collisions.
	Thus, the spectra of comets are marked by resonance fluorescence of the
	molecules excited by sunlight. 
	
	We observe only fragment species in the optical region of the spectrum. One
	of the most prominent set of molecular emissions observed are the 
	$\mathrm{d} ^3\Pi_g - \mathrm{a} ^3\Pi_u$ Swan bands of C$_{2}$.  The Swan
	bands are the dominant features in the green, orange and red part of
	the spectrum. Cometary spectra allow investigation of C$_2$ with longer path lengths and better vacuum than lab conditions. Thus, we are using observations of a comet to probe the structure of the C$_{2}$ spectrum and to determine the production pathways. 
	
	Outside the collisional zone, C$_2$ can reach high J levels because it has no pure rotational spectrum and only $\Delta\text{J}\pm1$, are allowed for electronic transitions (see Ref. \citenum{Herzberg}). The rotational temperature characterizes the distribution of J levels. \citeauthor{Lambert1990}\cite{Lambert1990} demonstrated that the observed C$_2$ Swan (0\,-\,0) band of comet 1P/Halley had two rotational temperatures, whereas their observations of an acetylene torch spectrum had only one. These two populations in Halley had temperatures, T$_\text{low} = 600-700$K  and T$_\text{high} \approx 3200$K. Previous rotational temperature calculations in comets for C$_2$ by \citeauthor{Lambert1983}\cite{Lambert1983}\,, \citeauthor{Ahearn1978}\cite{Ahearn1978}\,, and \citeauthor{Danylewych1978}\cite{Danylewych1978} only were modeled with a single temperature which was similar to the high temperature component of Halley. 
	
	In this paper, we present evidence for two rotational temperatures in comet 122P/de~Vico using modern spectroscopy modeling software and specifying the $\mathrm{d} ^3\Pi_g$ population function. This approach has several advantages, including simplicity, uniform treatment of uncontaminated blended lines, and the ability to treat perturbations. We also give estimates on some molecular constants for $\mathrm{b} ^3\Sigma^-_g (\nu = 10)$, which perturbs the $\mathrm{d} ^3\Pi_g \ (\nu = 0)$ state for N = 47.
	
	\section{Observations}
	We observed comet 122P/de~Vico using the 2.7\,m Harlan J. Smith Telescope of
	The University of Texas at Austin's McDonald Observatory. We used
	the high-spectral resolution Tull 2Dcoude spectrograph \cite{TullCoude}. 
	The spectra have a resolving power, R\,=\,$\lambda/\Delta\lambda$\,=\,60,000
	using a slit that was 1.2\,arcsec wide and 8.2\,arcsec long.
	Table~\ref{log} is a log of the observations.

	The comet was relatively bright, resulting in high signal/noise (S/N).  In
	addition, this comet has an extremely high gas-to-dust ratio, resulting in
	very little continuum from solar photons reflected off the dust. With so little dust, we neglected any solar spectrum removal since it plays such a small part in these observations. The spectral dispersion of the observations was computed using separate spectra of a ThAr lamp and had an rms error of the wavelength of $\sim24$\,m\AA. The spectra were Doppler shifted to the laboratory rest frame using the geocentric radial velocity of the comet.
	
	The $\Delta \nu=0$ band sequence of C$_{2}$ is spread over many echelle orders
	and each order has slightly different sensitivity.  We combined the
	orders together by correcting for the sensitivity variations using a
	separate solar spectrum obtained with the same instrument. This let us model the $\Delta\nu=0$ sequence from the (0\,-\,0) bandhead at 5165\,\AA \ to the (1\,-\,0) bandhead at 4737\,\AA \ with few strong contaminants. We chose to exclude the 4737-4783\,\AA \ section in the following analysis because it added only a few lines while adding many more noisy data points, thereby degrading the quality of the fit from imperfect noise removal.

	\section{Methods}
	We modeled the de~Vico data using the PGOPHER\cite{Colin2017} molecular modeling code.  PGOPHER is a code to model rotational, vibrational and electronic spectra of molecules. It can be used to fit a spectrum of line wavelengths and intensities or, as we used it, to input molecular constants to simulate a molecular spectrum. With its high degree of precision and customization, PGOPHER was well suited for our analysis. Serendipitously, the $\Delta\nu = 0$ band sequence up to (4\,-\,4) was already included in a sample constants file making it easy to start our analysis. \citeauthor{Brooke2013}\cite{Brooke2013} modeled up to (9\,-\,9), so we added in their constants to those that came with PGOPHER and have included vibrational levels (0\,-\,0) through (9\,-\,9) in our models.
	
	Unlike in a laboratory, where a single type of molecule is being studied, the comet spectrum consists of many different molecular emissions (see \citenum{devicoAtlas}).  Thus, the spectral region of the Swan bands can contain contaminants from other species, such as NH$_2$. In order that the contaminating lines not influence the quality of the fit, strong contaminating lines were masked out. Many features that were likely perturbed were also masked out. We further excluded wavelengths less than 4986.2\,\AA\ because of persistent offsets between the peaks of the model and data of 0.05 to 0.1\,\AA. This left 4107 data points covering the wavelength region 4986-5169\,\AA\ to model.

	\citeauthor{Lambert1990}\cite{Lambert1990} concluded that the Swan band population of comet Halley represented two different rotational temperatures. It is not clear how universal is the need in cometary spectra for populations at two temperatures. So we tested this by modeling our de~Vico spectrum both with a single temperature and with two temperatures.  
	The single temperature version used the built-in simulation temperature. The two-temperature distribution, $f$, had the form
	\begin{equation}
	f(a,T_1,T_2) = e^{(-E/(k_BT_1))} + ae^{(-E/(k_BT_2))}
	\end{equation}
	where $a$ is the ratio of the contribution of each population, and $T_1, T_2$ are the two temperatures. In addition, a data scaling factor, a$_\text{spec}$, was used for both fits since our data are not flux calibrated. a$_\text{spec}$ does not change the profile of the modeled spectrum. For both of the rotational temperature distributions, we also included models with and without a vibrational temperature (T$_\text{vib}$).

	The intensity error is dominated by Poisson noise from the CCD, which is proportional to $\sqrt{\text{Counts}}$ measured, so we use this as a first order weighting scheme. To include these weights in our fit, we made minor modifications to PGOPHER that have been incorporated into the current development version. By including weights, the influence of noise and weak features is reduced. 
	
	We also made a preliminary model for the N = 47 perturbation of d$^3\Pi_g(\nu=0)$ by b$^3\Sigma^-_g(\nu=10)$. While this perturbation has been suspected since \citeyear{Callomon1963}\cite{Callomon1963}, we could not find the molecular constants required to model it. If unaccounted for, perturbations were  a large source of error in our residuals. We used constants from \citeauthor{Chen2015}\cite{Chen2015} as a starting point for b$^3\Sigma^-_g (\nu = 10)$ and adjusted constants until the model achieved a reasonable match with our observations and the identifications given in \citeauthor{Tanabashi2007}\cite{Tanabashi2007} 
	
	\section{Results and Discussion}
\begin{figure*}
		\includegraphics[scale=0.65]{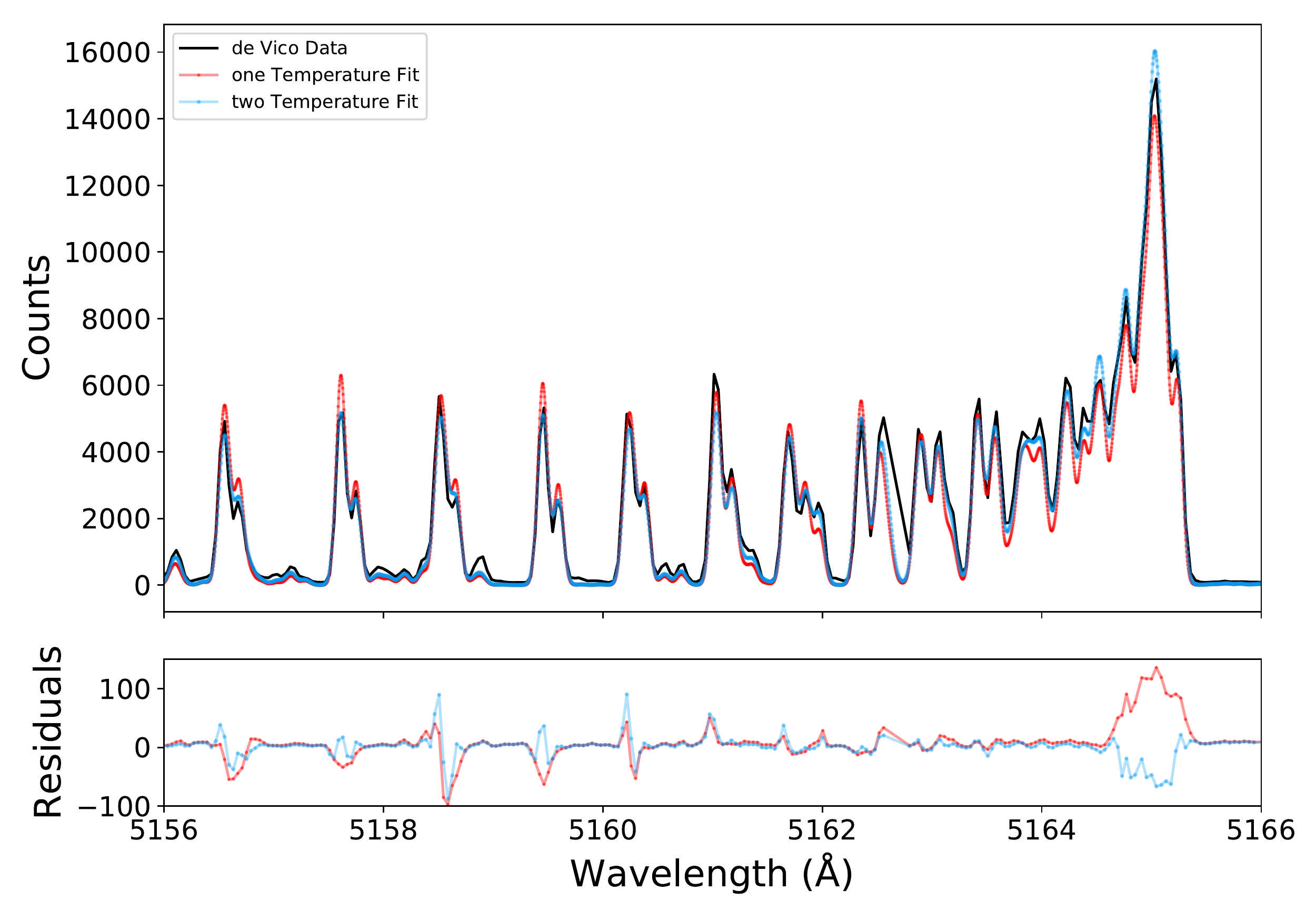}
		\caption{The de~Vico data near the (0\,-\,0) bandhead is shown along with the one- and two-temperature fits superimposed. While both models fit some parts of this spectral region well, the bandhead is fit better with the two-temperature model.}
		\label{fig:2}
	\end{figure*}

\begin{figure*}
		\includegraphics[scale=0.65]{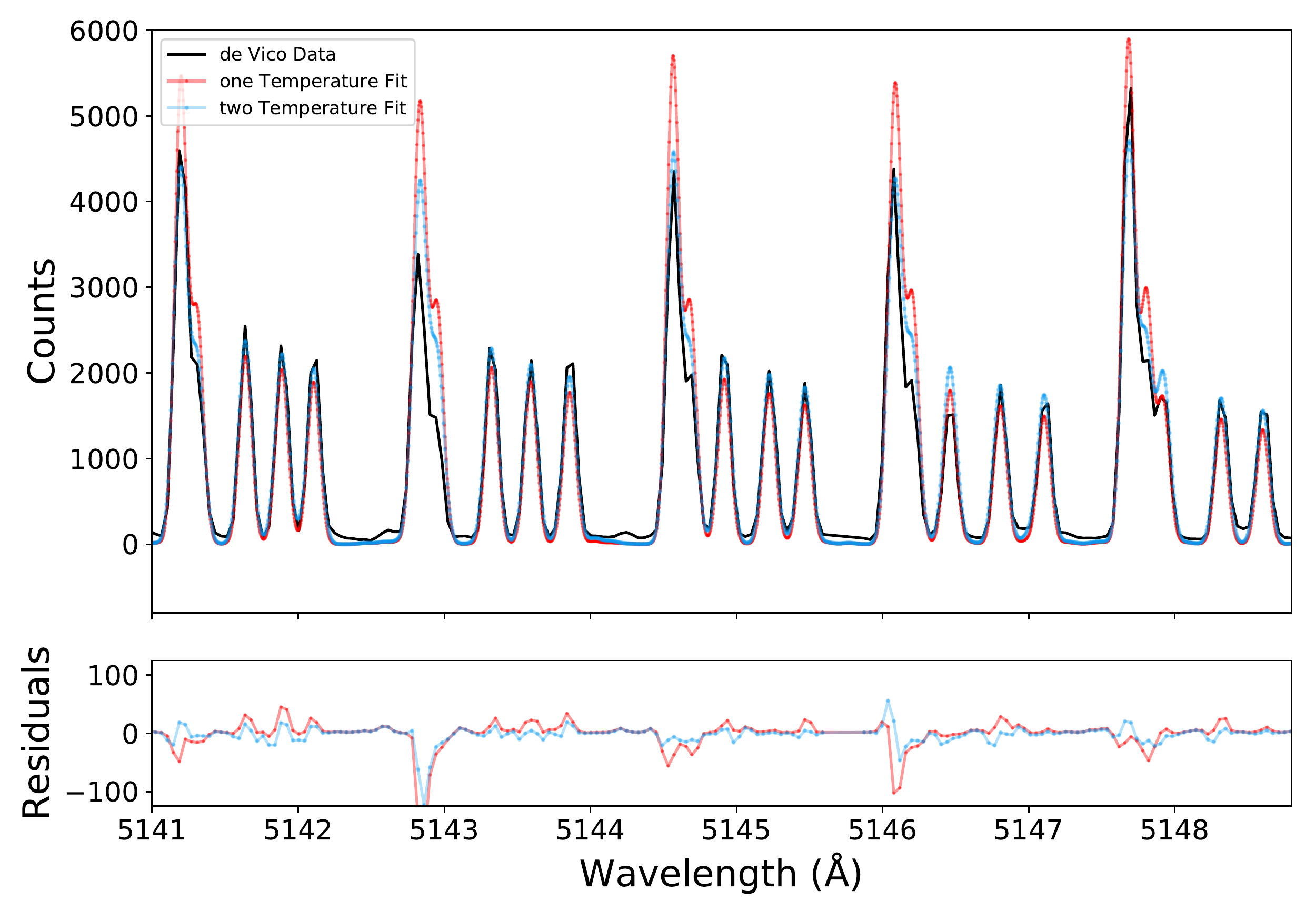}
		\caption{Comparison of the one- and two-temperature fits for some (0\,-\,0) P and R branch lines. The strongest lines are from the P branch, while the weaker lines are the R branch. Overall, the two-temperature model fits the data better than the one-temperature model. The P branch line at around 5143\,\AA\ is weaker than either model predicts. This might be the result of the Swings effect.}
		\label{fig:3}
	\end{figure*}

\begin{figure*}
		\centering
		\includegraphics[scale=0.65]{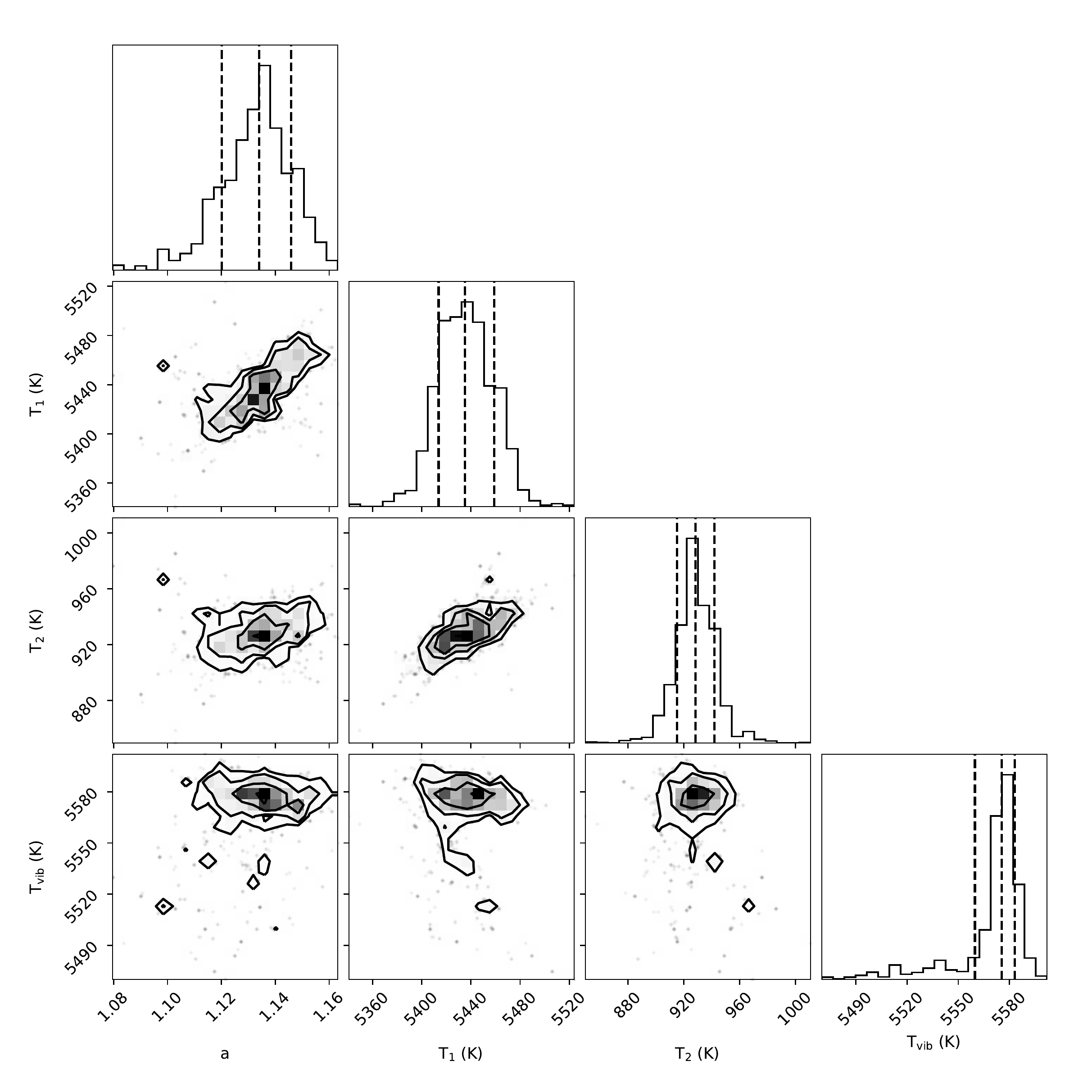}
		\caption{Posterior Probability distributions for two-temperature population. Contours are 1, 2, and 3$\sigma$, center dotted line on the histogram is the mean with the other dotted lines indicating the 1$\sigma$ confidence interval.}
		\label{fig:4}
	\end{figure*}

\begin{figure*}
	\centering
	\includegraphics[scale=.65]{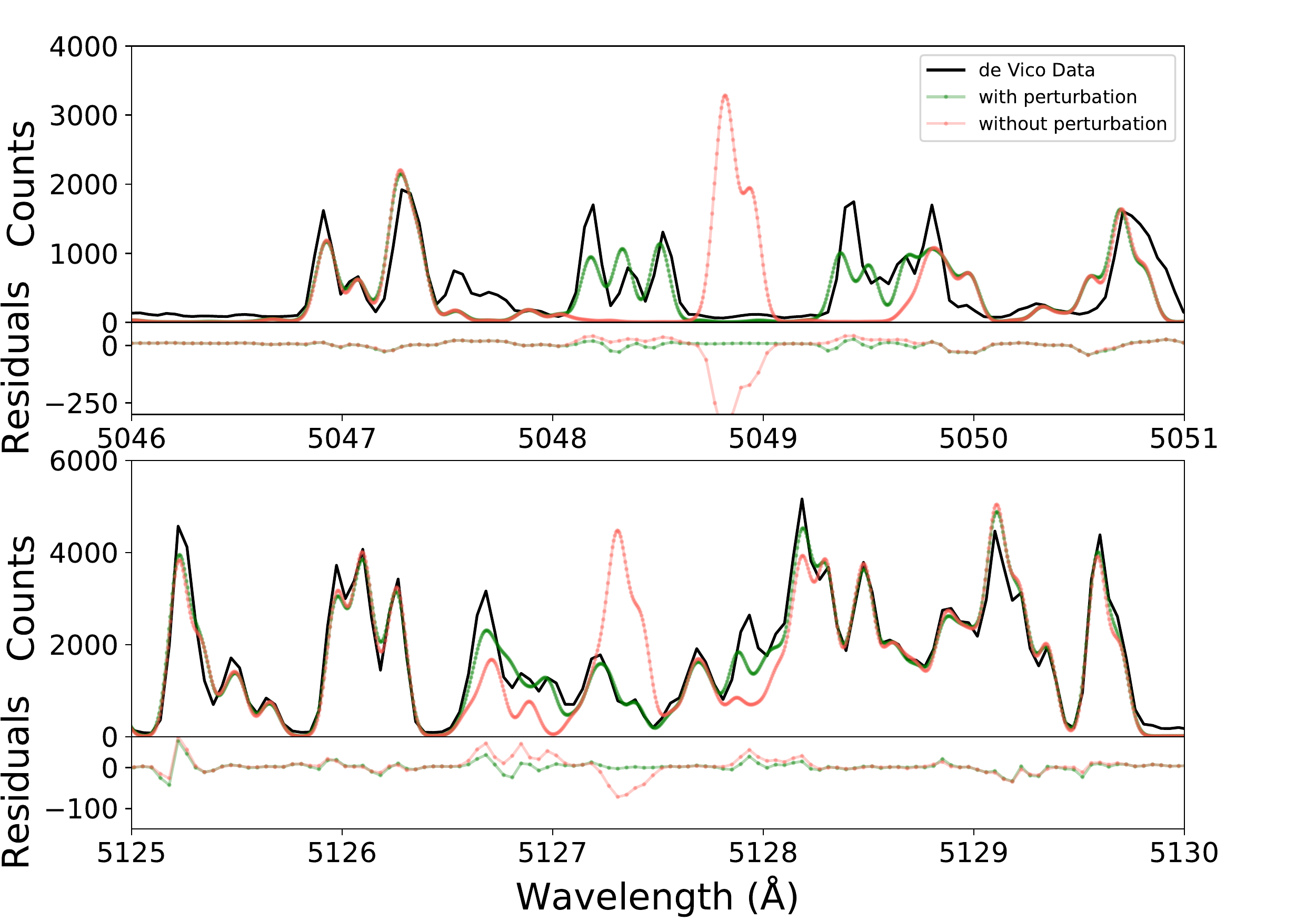}
	\caption{The two-temperature model is shown with and without the b$^3\Sigma^-_g (\nu = 10)$ perturbation included for the R branch (above) and P branch (below). The perturbed model removes the abnormally strong residuals at N = 47 seen in the unperturbed system while leaving the rest of the band undisturbed.}
	\label{fig:1}
	\end{figure*}

	The best fit values with 1$\sigma$ uncertainties and reduced chi-square ($\chi_r^2$) for the one- and two-temperature populations are given in Tables~\ref{tab:1} and \ref{tab:2} respectively. The instrumental broadening dominates the line width with a Gaussian $\text{FWHM} = 0.102$\,\AA. We convolved the models with this broadening to match the data. A model including all relevant perturbations is not available, so a minimum model involving just the states necessary was developed starting with the available constants\cite{Chen2015}\cite{Tanabashi2007} and adjusting as required. The constants used for the perturbing state are given in Table~\ref{tab:3}.  Visual comparisons of the one- and two-temperature models with the observations in a few important sections are given in Figures~\ref{fig:2} and \ref{fig:3}. The correlation matrix for the two-temperature fit is given in Table~\ref{tab:4}. We also used a Markov chain Monte Carlo (MCMC) model\cite{emceeDan} to examine the posterior distributions of the fitted parameters. The a$_\text{spec}$ parameter was found to be almost identical across all the fits, with a value of 5.26 adopted throughout. The posterior probability distributions for the two-temperature fit are given in Figure~\ref{fig:4}. The effect of the perturbation is illustrated in Figure~\ref{fig:1}. While the $\chi_r^2$ values are large for all of the populations used, there is an obvious preference for the two-temperature population. It is also clear that the contribution from the vibrational temperature cannot be excluded, as it improves models with both one and two rotational temperatures. Comparing the residual panels of Figures~\ref{fig:2} and \ref{fig:3}, it is apparent that the two-temperature model reproduces the (0\,-\,0) bandhead better than the single temperature. The one- and two-temperature models were similar for the (1\,-\,1) and (2\,-\,2) bandheads. We did not see bandheads for (3\,-\,3) or higher $\Delta\nu=0$ transitions. This is not surprising since the (2\,-\,2) bandhead is almost washed out by the (1\,-\,1) transition, and the (3\,-\,3) strength is less than half that of the (2\,-\,2). 
	
	We investigated the possibility of a three-rotational-temperature population. The fit with three temperatures was worse than that with two temperatures and actually converged toward the two-temperature solution by driving the coefficient of the third component to zero.
	
	The high value for $\chi^2_\text{r}$ that we derive probably comes from either errors in the reduction of the de~Vico data or unaccounted physics in the model that we have employed. However, inspection of the figures shows that the fits are quite good in general. Since de~Vico has high S/N and low dust, the most likely source of error in the data reduction is order de-tilting. The tilting is most pronounced toward the blue end of the spectrum, which is where the Swan band lines  are the weakest and have the least bearing on the quality of the fit. Therefore, we conclude that which physical processes are included/excluded drive our ``poor'' fit. Small differences in wavelength, usually less than 0.05\,\AA, between the peaks in the model and data were also seen. All models gave similar residuals around these mismatches so we conclude the temperatures are largely insensitive to this effect. The offsets could result from the peaks falling between two pixels on the CCD.  
	
\begin{figure*}
		\includegraphics[scale=0.65]{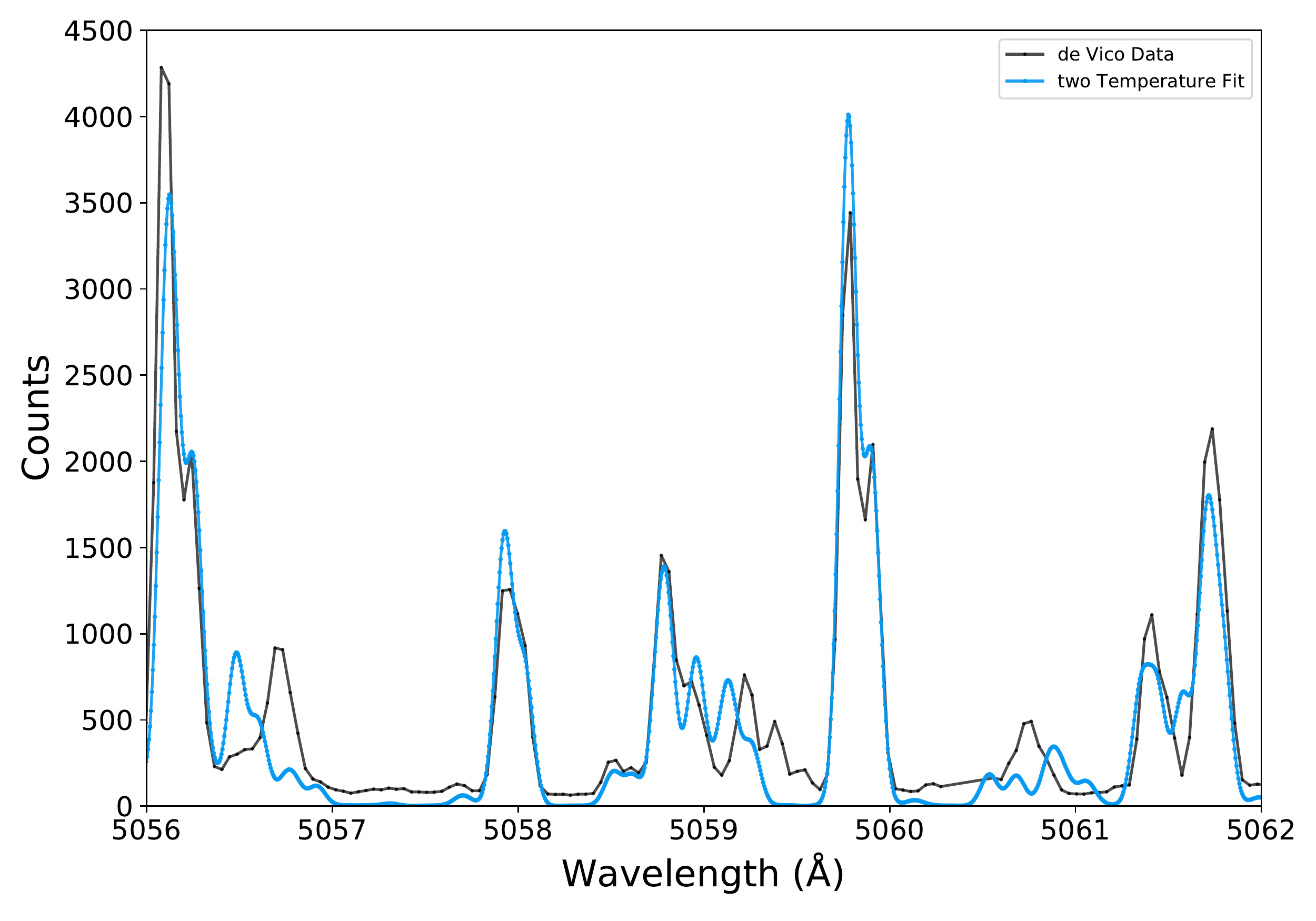}
		\caption{Examples of discrepancies between a two-temperature model and the observed spectrum are shown. Note the apparent shift of the (1\,-\,1) emission near 5056.5\,\AA.  It is likely that this shift is the result of perturbations that have not been included in the model.}
		\label{fig:5}
	\end{figure*}

	Inclusion of the d$^3\Pi_g \ \text{N}=47$ perturbation reduced the $\chi^2_r$ by more than 16\% in both models. This seems to be the strongest perturbation but is unlikely to be the only one. For example, the P branch of (1\,-\,1) N = 62 seems offset, as shown in the right panel of Figure~\ref{fig:5}. \citeauthor{Lambert1990}\cite{Lambert1990} saw intensity distributions of some P and R features that did not match the acetylene flame. If this mismatch is a consequence of the different environmental conditions then some deviation would be built in unless the different environments were dealt with explicitly. We removed blends with known contaminants to the best of our ability, but these frequently occur near or within C$_2$ lines, so a compromise must be reached so that not too much C$_2$ data is removed. These blends, especially for lines that were completely blended with C$_2$, will obviously enhance the observed intensity of the data with respect to the predictions. With many unidentified lines in the our bandpass, the possibility of undiscovered lines that are buried in strong C$_2$ emission could also exist. We also did not include the Swings\cite{Swings1941} effect, in which some lines that coincide with strong absorptions in the solar spectrum are suppressed since the C$_2$ band is produced via resonance fluorescence. 
	
	Our result agrees with previous work \cite{Lambert1990}. We also found that the two-temperature populations extend to the (1\,-\,1) and (2\,-\,2) bandheads. There are two explanations for this bimodal temperature. \citeauthor{Jackson1996}\cite{Jackson1996} found this as an outcome of the photolysis of C$_2$H, arguing that either the sudden or phase-space models of dissociation can apply. The sudden model is a classical treatment of the break up, where there is a maximum attainable rotational energy which corresponds to max J. In comparison, the phase-space model proceeds more slowly, allowing all J levels to be accessed. The sudden model gives a low J level distribution because the J values are constrained by classical conservation models, whereas the phase-space model allows for the high J level distribution. Which model applies depends on the trajectory of the H atom along the C$_2$H potential surface during photodissociation. Communication from Jackson to Lambert in 1989 (see \citenum{Lambert1990}) also indicated that a similar bimodal population can result from other parents/grandparents. The other explanation outlined by \citeauthor{Lambert1990}\cite{Lambert1990} is driven by intercombinational cooling from c$^3\Sigma^+_u - \text{X} ^1\Sigma^+_g$ and a$^3\Pi_u - \text{X}^1\Sigma^+_g$ transitions. Since the triplet-singlet transitions occur more readily than the singlet-triplet ones there is a net loss of energy from the triplet system. Thus a-X and c-X cool the Swan system. Simulation results by \citeauthor{Gredel1989}\cite{Gredel1989} and \citeauthor{KSwamy1997}\cite{KSwamy1997} offer support for this. 
	
	Quantifying the importance of the formation pathways versus the intercombinational transitions depends on our ability to distinguish or isolate them. A first attempt could be examining the spectrum as a function of cometocentric distance, assuming that the formation
	influence diminishes far away from the nucleus. This relies on the time for a C$_2$ molecule to establish fluorescence equilibrium with the Sun, $\tau_\text{eq}$. \citeauthor{Lambert1983}\cite{Lambert1983} and \citeauthor{Odell1988}\cite{Odell1988} estimated this value as $\tau_\text{eq} < 500$ seconds.  The lifetime against photodisocciation of C$_{2}$ at de~Vico’s heliocentric distance can be computed using the cometary scale lengths of Cochran\cite{Cochran1985} of $5.7 \times 10^4$ km and an outflow velocity of 1 km\,sec$^{-1}$.  Thus the expected lifetime of C$_{2}$ is of order $5.7 \times 10^4$  sec. Once appreciable amounts of new C$_{2}$ are no longer made, the contribution from formation should be small.
	
	\citeauthor{Jackson1996}\cite{Jackson1996} produced bimodal temperatures for the X$^1\Sigma^+_g$, A$^1\Pi_u$, and B'$^1\Sigma^+_g$ singlet states in the laboratory. They found each electronic state has significantly different temperatures. \citeauthor{KSwamy1997}\cite{KSwamy1997} predicted that the Mulliken and Phillips systems should also have bimodal temperatures. We intend to investigate whether there are two temperatures in the Phillips system for de~Vico observations in a later analysis.
	
	\citeauthor{KSwamy1997}\cite{KSwamy1997} seemingly reproduced the two-temperature distribution observed in comet Halley by including Phillips, Mulliken, Ballik-Ramsay, Swan, and the a$^3\Pi_u - \text{X}^1\Sigma^+_g$ systems. He assumed all molecules start in the $\nu = \text{J} = 0$ ground state and modeled them by exposure to a smoothed solar spectrum. He did not discuss why the photolysis of C$_2$H would produce this population. Applying either of the aforementioned photodissociation mechanisms implies some initial J distribution. What results from populations with other rotational, vibrational, and electronic levels that emulate cometary conditions is unknown. Krishna Swamy recovered a bimodal distribution without including the c-X transition. \citeauthor{Gredel1989}\cite{Gredel1989} found the rotational temperature strongly depends on both a-X and c-X strengths since both move energy into the singlet system. A simulation, testing different source functions, pressures, and initial population distributions would be of great interest to constrain what the bimodal temperature depends on.
	
	\section{Conclusions}
	In this paper, we have applied new models of the C$_2$ Swan band in comet 122P/de~Vico. These models fit most of the $\Delta \nu = 0$ band well by incorporating a two-temperature source population as well as including some perturbations between states.  We showed that this two-temperature model fits the data better than either a single rotational temperature or three rotational temperatures.  The reason for the two temperature populations is not entirely clear, though we quote several papers with potential models.  We plan to add to our understanding of the processes that produce the C$_2$ spectrum by using additional observations of comets. We will explore how universal is the two-temperature population with heliocentric distance and cometary orbital dynamical type in future work.
	
	\begin{acknowledgement}
		This work was performed under NASA grant NNX17A186G.  TN was supported by the Dean's Excellence Fellowship. 
	\end{acknowledgement}
	\providecommand{\latin}[1]{#1}
	\makeatletter
	\providecommand{\doi}
	{\begingroup\let\do\@makeother\dospecials
		\catcode`\{=1 \catcode`\}=2 \doi@aux}
	\providecommand{\doi@aux}[1]{\endgroup\texttt{#1}}
	\makeatother
	\providecommand*\mcitethebibliography{\thebibliography}
	\csname @ifundefined\endcsname{endmcitethebibliography}
	{\let\endmcitethebibliography\endthebibliography}{}
	
	\newpage

\begin{table*}
\begin{tabular}{l | c | c}

\multicolumn{1}{c}{Date (UT)} & Heliocentric & Geocentric \\
 & Distance (AU) & Distance (AU) \\
\hline
03 Oct 1995 & 0.66 & 1.00 \\
04 Oct 1995 & 0.66 & 0.99 \\
\hline
\multicolumn{3}{l}{Perihelion was 06 Oct 1995.}
\end{tabular}
\caption{Log of Observations}\label{log}
\end{table*}

	\begin{table*}
		\begin{tabular}{c | c | c | c}
			& \multicolumn{2}{c |}{With Perturbation} & Without Perturbation\\
			& Without T$_{\text{vib}}$ & With T$_\text{vib}$  & With T$_\text{vib}$\\
			\hline
			T$_\text{rot}$(K) & $5075\pm17$ & $4031\pm25$ & $3880\pm30$\\
			T$_\text{vib}$(K) & - & $5860\pm29$ & $6100\pm34$ \\
			$\chi_r^2$ & $560$ & $406$ & $492$\\
		\end{tabular}
		\caption{Single Temperature Fit Summary}
		\label{tab:1}
	\end{table*}
	
	\begin{table*}
		\begin{tabular}{c | c | c | c}
			& \multicolumn{2}{c |}{With Perturbation} & Without Perturbation\\
			& Without T$_{\text{vib}}$ & With T$_\text{vib}$  & With T$_\text{vib}$\\
			\hline
			a & $1.249\pm0.034$ & $1.112\pm0.037$ & $1.313\pm0.054$\\
			T$_1$(K) & $6130\pm45$ & $5432\pm76$ & $5663\pm103$ \\
			T$_2$(K) & $1179\pm43$ & $931\pm43$ & $1095\pm50$ \\
			T$_\text{vib}$(K) & - & $5519\pm25$ &  $5674\pm30$\\
			$\chi_r^2$ & $313$ & $280$ & $360$\\ 
		\end{tabular}
		\caption{Two Temperature Fit Summary}
		\label{tab:2}
	\end{table*}
	
	\begin{table*}
		\begin{tabular}{c | c}
			Origin & $19836.25468$ \\
			B & $1.5418643$ \\
			$\lambda$ & $0.3593$\\
			o & $-1.498$\textsuperscript{\emph{a}} \\
			$\gamma \times 10^4$ & $3.929$\\
			p$\times 10^5$ & $2.1$\textsuperscript{\emph{a}} \\
			D$\times 10^6$ & $6.417$\\
			H$\times 10^{12}$ & $4.88$\textsuperscript{\emph{a}} \\
			Strength & $0.03392826$\\
			\hline
			\multicolumn{2}{l}{}\\[-10pt]
			\multicolumn{2}{l}{\emph{a}: Values from \citeauthor{Chen2015}}
		\end{tabular}
		\caption{Molecular Constants for b$^3\Sigma_g^-(\nu=10)$}
		\label{tab:3}
	\end{table*}
	\begin{table*}
		\begin{tabular}{c | c | c | c | c }
			& a & T$_1$ & T$_2$ & T$_\text{vib}$\\
			\hline
			a & 1.0 & 0.779 & 0.552 & -0.290 \\
			\hline
			T$_1$ & 0.779 & 1.0 & 0.724 & -0.478 \\
			\hline
			T$_2$ & 0.552 & 0.724 & 1.0 & -0.224 \\
			\hline
			T$_\text{vib}$ & -0.290 & -0.478 &-0.224 & 1.0 \\
			\hline
		\end{tabular}
		\caption{Two Temperature Fit Correlation Matrix.}
		\label{tab:4}
	\end{table*}

	\newpage
	
	
	
\end{document}